\documentstyle[preprint,aps,epsfig]{revtex}

\begin{document}

\draft


\title{ 
       Comment on ``Comparison   of strangeness production \\ 
       between A+A and
       p+p reactions from 2 to 160 A GeV", \\
       by J. C. Dunlop and C. A. Ogilvie }

\author{Marek Ga\'zdzicki}

\address{CERN, Geneva, Switzerland, \\
         Institut f\"ur Kernphysik, Universit\"at Frankfurt, Frankfurt, Germany }

\author{Mark I. Gorenstein}

\address{Bogolubov Institute for Theoretical Physics, Kiev, Ukraine, \\
         Institut f\"ur Theoretische Physik, Universit\"at Frankfurt, Frankfurt, Germany }
 
\author{Dieter R\"ohrich}

\address{University of Bergen, 
         Bergen, Norway}

\date{\today}

\maketitle

\begin{abstract}

 A recent paper on energy dependence of 
 strangeness production in A+A and p+p interactions 
 written by  Dunlop and  Ogilvie 
 \cite{DO:00} indicates that there is a significant
 misunderstanding about the concept of strangeness enhancement
 and its role as a signal of Quark Gluon Plasma creation.
 In this comment we will try
 to clarify some essential points.
\end{abstract} 


\narrowtext

In a recent publication Dunlop and Ogilvie (DO)
\cite{DO:00} studied the energy dependence of strangeness
production and considered it from the point of view of a
possible transition to Quark Gluon Plasma (QGP).
In our opinion the DO paper does not reflect the 
status of the discussion on these subjects correctly 
and it is even  misleading.
We, therefore, feel the need to comment on the observation
of anomalous energy dependence of strangeness production
in nucleus--nucleus (A+A) collisions and its possible
interpretation.
In the first section of our comment we show that the data
analysis part of the DO paper repeats our earlier work
without, however, any reference to it.
In the second section
we argue that the  DO proposal 
to search for the transition by studying
the energy dependence of strangeness enhancement factor
is unjustified and misleading.

\noindent
{\bf I. Data on Strangeness Production}

In 1994  the idea that the creation 
of a transient QGP state
in the early stage of A+A 
collisions should be reflected in the collision
energy dependence of basic observables triggered
our analysis of results  
on hadron multiplicities
in A+A and nucleon--nucleon (N+N) collisions.
The work commenced from the relatively rich data on 
pion production \cite{pion}.
In 1996 this study was extended to 
strangeness production \cite{str}.
The basic experimental observations 
concerning strangeness 
here called strangeness anomaly
were the following:
\begin{itemize}
\item 
the strangeness to pion ratio in central 
A+A collisions increases
rapidly with the collision energy 
up to the top AGS (15 A$\cdot$GeV) energy; 
the ratio at top SPS (200 A$\cdot$GeV)
energy is approximately equal to the  
ratio at  AGS;
\item 
a very different energy dependence is observed for 
N+N interactions, a monotonic increase is
seen at all energies, between top AGS and SPS energies
the ratio increases by a factor of about 2.
\end{itemize}

This strangeness anomaly is rediscovered in the 
very recent DO paper \cite{DO:00}
on the basis of data on central Pb+Pb collisions
at SPS energy (158 A$\cdot$GeV)
\cite{NA49} and Au+Au collisions at
AGS energies \cite{E866}. 
After clarifying two points, the similarity becomes obvious:

1.) In the original paper \cite{str} the strangeness 
to pion ratio was
quantified by the measure 
\begin{equation}
E_S ~\equiv~
\frac{ \langle \Lambda \rangle + \langle K
+ \overline{K} \rangle }{\langle \pi \rangle}~,
\end{equation}
where  $\langle \Lambda \rangle$,
$\langle K + \overline{K} \rangle$ and
$\langle \pi \rangle$ 
are mean multiplicities of $\Lambda$ hyperons,
$K +  \overline{K}$ mesons and all pions, respectively.
For the comparison of the A+A collisions to nucleon-nucleon interactions
the N+N results were constructed by a proper
averaging of p+p, n+p and n+n data.
The above procedure yields the best experimental
estimate of the total strangeness to pion ratio. Our 
compilations indicate that the $K^+/\pi^+$ ratio
used in DO paper
is closely (within 20--30 \%) related to the $E_S$ measure.

2.) In refs. \cite{pion,str}
the energy dependence of pion and strangeness production
is presented by using the Fermi energy measure:
\begin{equation}
F~ \equiv ~(\sqrt{s} - 2 m)^{3/4}/\sqrt{s}^{1/4} 
~\cong~ \sqrt{s}^{1/2}~,
\end{equation}
where $\sqrt{s}$ is the c.m. energy for a 
nucleon--nucleon pair and
$m$ is the nucleon mass.
In the statistical model of the early stage of the collision
both entropy and the early stage temperature are approximately
proportional to $F$ \cite{early}.
In the DO paper the dependence on $\sqrt{s}$ is studied.

Keeping these two technical differences in mind one
observes that Figs. 1, 2, 3(5) and 4  in the DO paper
are almost identical
to, respectively, 
Figs. 3(4) in \cite{str}, 1 in \cite{pion},
7(8) in \cite{str} (for the most recent version see
also Fig. 3 in \cite{Ga:99}) and Fig. 3 in \cite{dr}.  
Therefore, the analysis presented in the DO paper
fully confirms 
the observations made in refs.~\cite{str,Ga:99,dr}
({
however, without
any references to these papers}).

\vspace{0.2cm}
\noindent
{\bf II. Interpretation of Strangeness Anomaly}

The anomalous energy dependence of the strangeness to
pion ratio in A+A collisions (see the previous
section)
was established experimentally and its connection with a possible
observation of transition to QGP in A+A collisions between 
AGS and SPS was stressed in Refs.~\cite{str,early}.
Note also that it serves as one of the basic arguments
for the current study of the low energy (40 A$\cdot$GeV)
Pb+Pb collisions at CERN SPS \cite{add-2}.
In contrast to this suggestion
it is proposed in the DO paper that in order to search
for the transition to QGP one should study the energy
dependence of the strangeness enhancement factor, i.e. 
the increase of the strangeness to pion ratio between
N+N and A+A collisions.
The idea of strangeness
enhancement as a QGP signal was formulated about 
20 years ago \cite{Ra:82}
and was based on the estimate that the strangeness 
equilibration time
in QGP is of the same order ($\approx 10$ fm/c) as
the expected life time of the fireball created in A+A
collisions.
Thus in the case of QGP formation the strangeness content is expected 
to approach its equilibrium value in QGP.
This equilibrium value is significantly higher (factor 2 or
more depending on energy) than the  strangeness
production in elementary N+N interactions.
Furthermore it was estimated that the
strangeness production in secondary hadronic interactions
which may follow initial N+N interactions
is small \cite{Ko:86}. 
Therefore, if QGP is not formed,
the strangeness yields would be 
expected to be much lower than those predicted by
equilibrium QGP  
calculations\footnote{In the DO discussion of the results
of the data analysis performed within hadron gas models the usage of the 
model parameters $\gamma_S$ (strangeness saturation
factor) and $\lambda_S$ (strangeness suppression factor)
have apparently been confused.}.
For experimental and theoretical reasons it is convenient to analyse
the strangeness to pion ratio in an actual study of strangeness production.
Then a simple and elegant signature of
QGP creation appeared:
a transition to QGP should 
be signalled by an increase of the strangeness to pion
ratio
from the value close to that measured in N+N interactions 
at the corresponding energy to the level  
given by QGP in equilibrium. 
This idea motivated study of the strangeness to pion ratio
in A+A collisions relatively to the corresponding ratio
in N+N interactions as a function of nuclear mass number and
collision energy. 

The comparison of the above expectations with the data
was for the first time possible in 1988 when the preliminary
results from S and Si beams at SPS and AGS were presented.
Experiment NA35 reported \cite{Ga:89} that in central S+S collisions
at 200 A$\cdot$GeV the strangeness to pion ratio is two times higher
than in N+N interactions at the same energy per nucleon.
Surprisingly
even larger enhancement (a factor of about 3)  was measured
by E802 in Si+A collisions at AGS \cite{Vi:89}. 
Recent data on central Au+Au collisions at low AGS energies
\cite{E866} complete the  picture:
strangeness enhancement is observed at all energies,
it is even stronger  at lower energies than at the SPS
energy. 
Thus the interpretation in line with
the original concept (strangeness enhancement $\rightarrow$ QGP)
was put in question from the very beginning by the AGS data. 
At the low AGS energies one does not expect the creation of QGP
and therefore one should not observe substantial strangeness
enhancement.
Consequently AGS measurements of strangeness enhancement larger
than that at SPS clearly leads to the
following conclusion:
{\it the simple concept
of strangeness enhancement as a signal of QGP is incorrect}.
Furthermore the observed enhancement is very large
(factor several), indicating that a potential connection
between strangeness measured in A+A collisions and that produced
in initial N+N interactions is lost.
Therefore, the proposal of the DO paper
to search for the transition to QGP by studying the energy
dependence of the 
strangeness enhancement factor
should be treated as unjustified and misleading,
it is not based on any specific model or at least an intuitive
physical picture. 

From our point of view it is more natural 
to concentrate on the strangeness to pion ratio
in A+A collisions instead of relating it to N+N interactions,
which may only introduce additional problems into the
data interpretation.
This suggestion is justified by
the statistical model of the early stage
of A+A collisions \cite{early} which, 
for a reasonable selection of the model parameters, 
predicts a transition to QGP between AGS and SPS
energies and explains 
the observed strangeness anomaly by the presence
of this transition.
We would like to stress once more the simplicity of the
argument, which relies on the basic assumption of the
model that in both confined and deconfined matter a maximum
entropy state is created at the early stage of the collisions. 
In confined matter the mass of strange and non--strange degrees
of freedom are expected to be of the order of
kaon and pion masses 
($m_S \cong 0.5$ GeV, $m_{NS} \cong 0.14$ GeV), respectively,
and the temperature is $T < T_C \cong 0.170$ GeV.
Thus, strange degrees of freedom are heavy ($m_S > T$)
and non-strange light ($m_{NS} < T$).
Therefore, in the statistical model one obtains approximately
\begin{equation}
\frac {\langle n_S \rangle} {\langle n_{NS} \rangle} ~ \sim ~
 \frac {T^{3/2}~ \exp(-m_S/T)} {T^3}~,
\end{equation}
which is an increasing function of $T$ for $T< T_C (<2m_S/3)$.
In QGP $m_S \cong 0.170$ GeV and $m_{NS} \cong 0.01$ GeV.
Both strange and non-strange degrees of freedom are light
($m \le T$), and therefore one gets approximately
\begin{equation}
\frac {\langle n_S \rangle} {\langle n_{NS} \rangle}~ \cong
~ {\rm const}(T)~ .
\end{equation}
Consequently
within the statistical model of the early stage we find
a simple explanation for a fast increase of
the strangeness to pion ratio at low energies (Eq. 3) and its 
rapid saturation
at high energies (Eq. 4) provided that 
the threshold for QGP creation is crossed.

Two striking predictions follow from our approach.\\
1.) {\it Very similar  strangeness to pion ratio
is predicted for SPS, RHIC and LHC energies.}
This follows from the observations that the model
predicts the transition to QGP to occur below top SPS
energy (and above top AGS energy) and that in QGP
the ratio $\langle n_S \rangle / \langle n_{NS} \rangle$ 
is almost independent of temperature (collision energy).\\
2.) {\it A non--monotonic (or kinky) energy dependence of the
strangeness to pion ratio is predicted in the vicinity
of the transition region.}
This is due to the fact that the model predicts the 
transition to occur at an energy where the 
strangeness to entropy (pion) ratio in the confined matter
is higher (or equal) than in the QGP.
Thus an initial fast increase of the ratio is expected to
be followed by a decrease (or a saturation) 
to the level expected in an equilibrium
QGP. The decrease starts at the onset of the transition
region ($\approx 30$ A$\cdot$GeV) and ends when the pure
QGP phase is produced ($\approx 60$ A$\cdot$GeV) \cite{early}.

In Fig. 1 the discussd above predictions  
of the statistical model of the early stage 
calculated for $T_C = 200$ MeV \cite{early} 
are confronted with the experimental
data on strangeness to pion ratio. 
In the case of $T_C = 170$ MeV the onset of transition 
takes place at the top AGS energy and the non--monotonic
behaviour is substituted by a kink in the energy dependence.
In general, one might try to extend the model calculations to N+N
interactions; in this case, however, canonical, and at low
energy even micro-canonical, calculations are needed.
Note that this introduces additional dependence on model
assumptions and  parameters and therefore also from the
point of view of our model study of strangeness enhancement factor
only confuses the interpretation of the data.

\vspace{0.3cm}
We hope to have shown that the idea to search for QGP by 
studying the strangeness enhancement factor, as proposed by DO, 
is unjustified and may lead to misleading conclusions.

\vspace{1cm}
\noindent
{\bf Acknowledgements}

We thank M. van Leeuwen for comments on the manuscript.
We acknowledge the
financial support of BMBF and  DFG, Germany.

\begin{figure}
\epsfig{figure=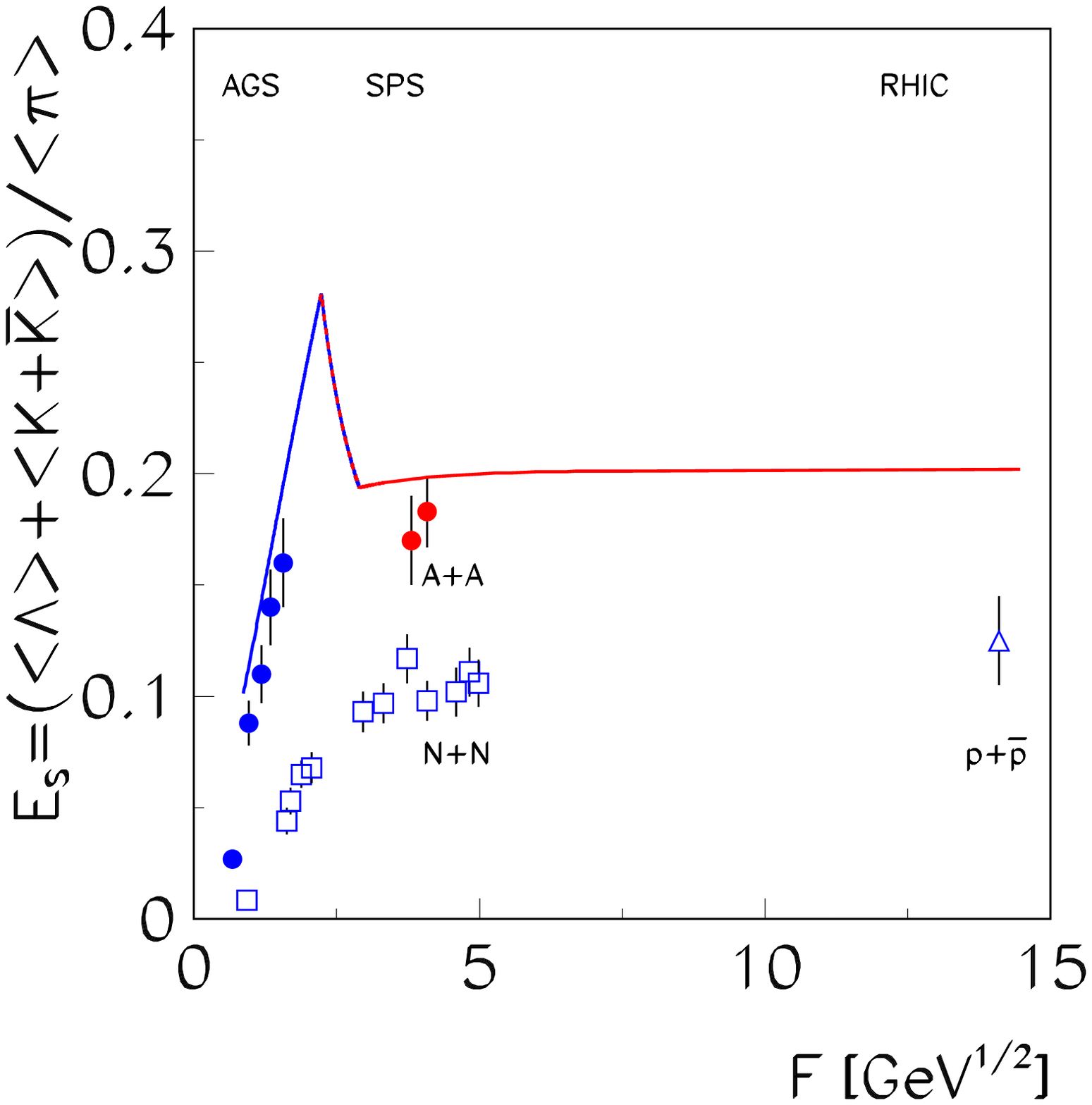,height=15cm}

\vspace{1cm}

\caption{Collision energy dependence of strangeness to pion ratio
for central nucleus--nucleus 
\protect\cite{aa} collisions (closed points),
nucleon--nucleon \protect\cite{str}
and $p+\bar{p}$ \protect\cite{ppbar}  interactions  (open points).
The predictions of the statistical model of the early stage
for A+A collisions are indicted by solid line. Within the model
one expects a
non--monotonic dependence of the strangeness to pion ratio
in the vicinity of the transition region
and its saturation at the value characteristic for equilibrium
QGP at high energies.}
\end{figure}

\end{document}